\documentstyle[preprint,revtex]{aps}
\begin{document}
\begin{title}
Unstable Periodic Orbits in the Stadium Billiard
\end{title}
\author
{Ofer Biham}
\begin{instit}
Department of Physics,
Syracuse University,
Syracuse, NY 13244
\end{instit}
\author
{Mark Kvale}
\begin{instit}
Laboratory of Atomic and Solid State Physics,
Cornell University,
Ithaca, NY 14853-2501
\end{instit}

\begin{abstract}
A systematic numerical technique for
the calculation of unstable periodic orbits in the
stadium billiard is presented.
All the periodic orbits up to order $p=11$ are
claculated and then used to
calculate the average Lyapunov exponent
and the topological entropy.
Applications to semiclassical quantization and to experiments in
mesoscopic systems and microwave cavities are noted.
\end{abstract}


\section{Introduction}
Chaotic billiards have attracted much interest in recent years
in the contexts of both classical chaos
\cite{sinai,bunimo1,bunimo2,benet}
and quantum chaos
\cite{mcdon,heller,bogo,sieber,tanner}.
The dynamics of a classical particle in various billiard geometries has
been studied extensively and the  mechanisms that produce
chaos in such systems have been examined and classified.
In particular, it was shown that for a class
of billiard geometries that includes the stadium billiard the dynamics is
ergodic
\cite{bunimo2}.
The quantum mechanics of billiards
which are chaotic in the classical limit has been studied.
The statistical properties of energy
spectra and the wave functions
have been analyzed in
an effort to characterize the quantum mechanical properties of such
systems.
Among the remarkable phenomena that have been observed are the
correspondence between the classical behavior and the
distribution of nearest neighbor energy level spacings and the
appearance of ``scars'' in the wave functions.

Unstable periodic orbits
provide a useful and systematic framework for
calculations both in the classical and the semiclassical regimes.
In  classical chaos these orbits can be used to calculate
fractal dimensions, entropies and Lyapunov exponents in a systematic
fashion
\cite{auer1,greb1,cvit1,cvit2,artuso}.
In the semiclassical regime they can be applied
to the calculation of the energy levels
and wave functions
using the Gutzwiller trace formula
\cite{bogo,gutz1,berry1,berry2,cvit3}.
These studies provide important insights into the
quantum mechanics of classically chaotic systems.
They are also relevant to experiments on billiard
shaped microwave cavities
\cite{stock,srid},
to mesoscopic systems in the ballistic regime
\cite{mesos}
and to a variety of scattering problems
\cite{bs}.

A systematic method for the calculation of unstable periodic
orbits in discrete dissipative
maps has been proposed recently and applied
to the H\'enon map
\cite{bw1}
and the dissipative standard map
\cite{wb}.
In this method one first identifies a symbolic dynamics for the
system such that each periodic orbit is associated with a
unique symbol sequence. Then the periodic orbits can be calculated
systematically by picking each symbol sequence and calculating the
orbit associated with it.
This way all the periodic orbits up to a given order can be obtained
to a very good accuracy.

In this paper we extend the method and apply it to the
calculation of unstable periodic orbits in the
stadium billiard.
We identify a useful symbolic dynamics for the stadium
and verify its uniqueness numerically, using the results
to calculate the topological entropy and the average Lyapunov exponent.
The paper is organized as follows. In Section II we present the method
for calculation of unstable periodic orbits. We test the method in
Section III and apply it to the calculation of average Lyapunov
exponent and the topological entropy. The results are summarized
and discussed in Section IV. In the Appendix we provide a list of
the geometrical pruning rules.

\section{The Method}

In this Section we describe the  numerical technique for the
calculation of unstable periodic orbits in chaotic billiards.
We apply it to the stadium billiard and calculate all
periodic orbits up to order 11 and selected orbits up to order
200.
These results allow us to perform a systematic calculation of
the average Lyapunov exponent and the topological entropy.

Consider the dynamics of a classical particle
in the stadium billiard where the potential
is zero inside and infinity outside.
The trajectory of the particle is a straight
path which is broken and reflected whenever the particle
hits a boundary.
It can be described by
a map which gives the functional relation between the
position and orientation of two successive bounces of the
particle from the walls.
A periodic orbit in the stadium billiard is shown in Fig. 1.
In general, a periodic orbit of order $p$ can be described
by a sequence of the form
$(S_1,x_1),\dots,(S_p,x_p)$
where
$(S_{p+1},x_{p+1})=(S_1,x_1)$,
$S_n$
specifies the boundary segment that the particle hits
and
$x_n$
specifies the position on that particular segment.
The symbol $S_n$ can take one of four values 0, 1, 2 or 3 as
shown in Fig. 1.
On a straight boundary $-a<x_n<a$
(where $a$ is the half length of the straight edge),
while on a circular boundary
$-\pi/2<x_n<\pi/2$ represents an angle.
Note that $0<a<\infty$ is the only parameter that determines
the shape of the stadium.
In the analysis that follows we find it useful to
continue the straight boundaries to infinity from
both sides and to complete the circles (thin lines in Fig. 1).

We will first try to obtain the
symbolic dynamics of the stadium
billiard.
Let us examine the set of symbol
sequences $S_1,\dots,S_p$ to see whether they provide a
unique identification of the periodic orbits.
We find that this symbolic representation is unique
with one exception:
in case of two or more successive bounces from one semicircle
there is a degeneracy between the upward and downward trajectories
(see Fig. 2), namely there may be two periodic orbits
with the same symbol sequence.
To remove this degeneracy we introduce the  symbols
$1^{\uparrow}$,  $1^{\downarrow}$
and
$3^{\uparrow}$,  $3^{\downarrow}$
which replace the symbols 1 and 3 respectively,
such that the arrows identify the direction of the
motion
(note that for a single bounce from a semicircle there is no such
degeneracy. Therefore, in this case only one symbol is needed for
each semicircle and we choose $1^{\uparrow}$ and $3^{\uparrow}$
for the semicircles on the left and right hand sides respectively).
We now have six symbols:
$0$, $1^{\uparrow}$, $2$, $3^{\uparrow}$,
$1^{\downarrow}$, and $3^{\downarrow}$,
that provide a unique symbolic dynamics,
namely that each periodic orbit corresponds to a unique
symbol sequence.
In order to perform a systematic study of all the periodic
orbits of a given order $p$ it is convenient to represent
the symbol sequences as numbers in base 6.
We thus replace the symbols
$0$, $1^{\uparrow}$, $2$, $3^{\uparrow}$,
$1^{\downarrow}$, and $3^{\downarrow}$
by  0, 1, 2, 3, 4, and 5 respectively.

We have obtained strong numerical evidence that this
symbolic representation is unique, namely each symbol
sequence cannot represent more than one orbit
\cite{twocyc}.
On the other hand, some of these sequences do not represent
periodic orbits of the stadium billiard and should thus be pruned.
There are two types of pruning rules:
(a) {\it geometric} pruning rules,
that prune sequences that cannot,
in principle,
represent any orbit
no matter what the value of $a$ is
(the simplest example is a sequence that has
two or more successive bounces from a straight edge)
(see Appendix A);
(b) {\it dynamical} pruning rules,
that prune periodic orbits which
can exist in principle but do not exist for a given value
of the parameter $a$,
because they cannot satisfy the reflection condition
on the boundaries of the stadium.

We will now describe the computational technique for the
calculation of the periodic orbits.
It is based on the idea that a periodic orbit of order $p$
can be obtained from a variational
function of $p$ variables.
We visualize the periodic orbits as a set of interacting atoms
sitting on the boundaries of the stadium.
One can write a variational function which is minimized
when the reflection condition between the incoming and outgoing
directions is satisfied at each point.
The variational function is
\begin{equation}
V = \sum_{n=1}^p
(  \cos \theta_n^+ - \cos \theta_n^- )^2
\end{equation}
where $\theta_n^-$
$(\theta_n^+)$
is the angle between the incoming (outgoing) direction and the
tangent at the reflection point.
We will now
introduce an
artificial dynamics that converges to the minimum
of this variational function and thus allows us to
find the periodic orbits.

We apply the following procedure:
(a) choose a symbol sequence
$S_1,\dots,S_p$, $S_n=0,1,\dots,5$
that is not
forbidden by the geometric pruning rules;
(b) choose an initial condition
$x_1,\dots,x_p$
where each atom sits on the correct boundary section defined
by the symbol sequence, but in an arbitrary position;
(c) introduce an artificial dynamics
that converges to the desired periodic orbit identified by
$S_1,\dots,S_p$.
The artificial dynamics takes the following form:
\begin{equation}
{dx_n \over dt} = C_n \cdot \vert
 \cos \theta_n^+ - \cos \theta_n^- \vert ,
n=1,\dots,p
\label{artdyn}
\end{equation}
where the right hand side represents the force applied on the
nth atom and all these equations are iterated simultaneously.
The prefactor is $C_n=\pm C$ where $C>0$ is a constant
(typically $C=1$)
and the signs are chosen such that
on a straight edge the atoms move to the
direction of the smaller angle, while on a semicircle
they move to the direction of the larger angle
(see Fig. 3).
In both cases the smaller angle then increases and
the larger angle decreases until they become equal, so
the reflection condition is satisfied and the periodic
orbit is obtained
\cite{directions}.
Note that this dynamics preserves the topological structure
of the orbits, i.e., the symbol sequence never changes.
We find that for any symbol sequence which is not forbidden by the
geometric pruning rules
this artificial dynamics converges to
a fixed point.
In case that all the reflection points lie on boundary segments
of the original stadium then the resulting configuration is
a periodic orbit of the stadium billiard.
However,
for some choices of $S_1,\dots,S_p$
the resulting orbit has one or more atoms
on boundary segments which do not belong
to the original stadium ( thin lines in Fig. 1).
These sequences do not represent periodic orbits of the
stadium and are thus dynamically pruned.

In practice we solve Eqs.
(\ref{artdyn})
until all forces become smaller than a test value $\epsilon$
(typically $\epsilon=10^{-7}$).
The procedure converges for all initial conditions.
Since only the final configuration is of interest it is
possible to choose a fourth order Runge-Kutta method
with a relatively large step size ($h=0.1$) to solve
the equations.

\section{Results}

To examine the technique we have calculated all the periodic orbits
up to order $p=11$ and selected orbits up to order 200
\cite{shrt}.
An important test is the uniqueness of the symbolic dynamics.
The symbolic dynamics is unique if:
(a) each periodic orbit corresponds to
one and only one symbol sequence;
and
(b)  each symbol sequence corresponds to at most one periodic orbit.
The symbolic dynamics we introduce for the stadium
guarantees property (a).
This is due to the fact that the symbol sequence is determined by the
sequence of boundary segments on which the atoms sit.
This sequence is determined in the initial conditions and does not
change during the iterations of the artificial dynamics
(note that in other systems such as
the H\'enon map this property is not guaranteed as in principle
the same periodic orbit can be obtained for two different symbol
sequences).
To examine property (b) we repeated the calculation of periodic
orbits with different sets of initial conditions. In all cases we
found for any choice   $S_1,\dots,S_p$ the artificial dynamics (2)
converges to a unique orbit independent of the initial conditions.
Therefore we have strong numerical evidence for the uniqueness of
our symbolic dynamics.

Using our data up to $p=11$
we then calculated the topological entropy, which is the limit of
$K_0 (p)=1/p \cdot ln N(p)$ as $p \rightarrow \infty$
(see Table 1).
Here  $N(p)$  is the total number of points which belong to
periodic orbits of order p
\cite{nopo}.
Note that time reversal orbits need to be included
(see Fig. 4).
The topological entropy as a function of p is shown in
Fig. 5.
To calculate the Lyapunov number for a periodic orbit we use the
monodromy matrix method.
For each path, we define the
the infinitesimal angular deviation
$d \theta$,
and  the normal deviation $d y$ which are
canonically conjugate variables.
Then the Lyapunov number of a periodic orbit of order p
is obtained from
the matrix
$M = \prod_{i=1}^p M_i$,
where $M_i$ takes the form

\begin{eqnarray}
            M^s_i = \left(\matrix{1&0\cr-l_{i-1}&1\cr}\right) \qquad
\end{eqnarray}

for a point on a straight edge, and

\begin{eqnarray}
            M^c_i = \left(\matrix{(2\alpha_i l_{i-1}/R)-1&2\alpha_i/R\cr
             -l_i&-1\cr}\right)
\end{eqnarray}
for a point on a semicircle.
Here $l_i$ is the path length from atom $(i-1)$
to atom $i$, $\alpha_i=1 / \sin \theta_i$
(where  $\theta_i=\theta_i^+=\theta_i^-$ )
and $R$ is the
radius of the circle (we use $R=1$).
The Lyapunov number $\Lambda$
is then the largest eigenvalue of
the matrix $M$
and the corresponding Lyapunov exponent is
$\lambda=ln \Lambda$.

The periodic orbits can then be used to calculate average properties of
the chaotic dynamics.
The motion of a particle in the stadium can be considered as
a continuous motion, for which the pointwise Lyapunov exponent
is $\lambda_c=\lambda / L$
where L is the length of the orbit.
It can also be considered as
a discrete map of the form
\begin{equation}
(S_{n+1},x_{n+1},\theta_{n+1}^+)=F(S_{n},x_{n},\theta_{n}^+)
\end{equation}
that relates the coordinates of consecutive
bounces from the boundaries.
In this case
the pointwise Lyapunov
exponent is
$\lambda_d=\lambda / p$, where $p$ is the number of boundary hits.
Considering the motion as a discrete map we can
use the periodic orbits to
calculate the
average discrete Lyapunov exponent given by
the limit as $p \rightarrow \infty$ of
\cite{greb1}:
\begin{equation}
\langle \lambda_d \rangle =
 {1 \over p}
 { { \sum { \lambda } e^{-\lambda} }
\over
{    \sum e^{-\lambda} } }.
\label{liapdis}
\end{equation}
Here the sum goes through all the
fixed points of $F^p$. These include all the points that belong to
periodic orbits of order p and its divisors.
The Lyapunov exponents $\lambda$ are the Lyapunov exponents of
the periodic orbits of order $p$ including repetitions of shorter
orbits.
The average continuous Lyapunov exponent is calculated in
similar fashion and is given by:
\begin{equation}
\langle \lambda_c \rangle =
 { { \sum { \lambda \over L } e^{-\lambda} }
\over
{    \sum e^{-\lambda} } }
\label{liapcon}
\end{equation}
where $L$ is the length of the orbit.
The average discrete and continuous Lyapunov exponents
as a function of p are
shown in Fig. 6.
We find that the convergence as a function of $p$ is slow.
There are oscillations between even and odd $p$ such that
the average Lyapunov exponents are larger for odd $p$ and
smaller for even $p$. We observe that better convergence
can be obtained by using the average of
$ \langle \lambda_d \rangle $
(and $ \langle \lambda_c \rangle $)
between pairs of consecutive even and odd values of $p$.
We then find that $\lambda_c=0.43 \pm 0.02$ in agreement
with previous results
\cite{benet}
which are based on the iteration of an ensemble of long
trajectories with random initial conditions.

\section{Summary}

In summary, we have developed a numerical technique for the
calculation of unstable periodic orbits in the stadium billiard
which provides a complete classification of all periodic orbits
and allows us to calculate any particular orbit to
a very good accuracy.
Using this technique
we found all
the periodic orbits up to order 11 and used them to calculate
quantities that characterize the dynamics
such as the topological entropy and the average
Lyapunov exponent.
We have also conducted
preliminary studies with the Gutzwiller trace formula
that will be reported in a future publication.
We are currently considering a generalization of the method
to the case where magnetic field is present and the orbits consist of
circular segments.
We believe that similar concepts can be developed for other billiard
geometries of interest and scattering problems.
Such techniques will be useful
to study the dynamics of electrons in mesoscopic
systems in the ballistic limit
\cite{mesos}
and to a variety of scattering problems
\cite{cvit3,bs}.

\acknowledgements

We thank S. Fishman, C. Jayaprakash
and W. Wenzel for helpful discussions.
We acknowledge support from the National Science Foundation
under Grant No.
DMR-8451911,
an IBM Post-doctoral fellowship (MK) and
the Department of Energy
at the Ohio State University
where this work was initiated.
This work was conducted using the computational resources of
the Ohio Supercomputer Center at the Ohio State University
and the Northeast Parallel Architectures Center (NPAC) at Syracuse
University.
We thank the referees who reviewed a previous version of this paper
for useful suggestions. While revising the paper we received
a preprint from Hansen
\cite{hansen}
using our symbolic dynamics to calculate pruning fronts.

\appendix{The Geometric Pruning Rules}

In this Appendix we present the geometrical pruning rules, namely
the pruning rules that hold for all values of the parameter $a$.
These rules that result from the geometry of the stadium billiard
were found by inspection and are easy to verify.
The following symbol sequences are disallowed due to geometrical
pruning rules:

\noindent
(1) Sequences that contain two consecutive reflections from the same
straight boundary ($...00...$ or $...22...$, where $...$ represents
any set of symbols from 0 to 5).

\noindent
(2) Sequences that contain an isolated 4 or an isolated 5.
For the sake of
uniqueness a single bounce from a circular boundary is represented by
1 or 3, and therefore 4 and 5 appear at least in pairs (...44... or
...55...).

\noindent
(3) Each periodic orbit (except for the sequence $02$ that represents
a continuous family of period two orbits that bounce between the
straight boundaries) should hit each one of the circular boundaries at
least once. Therefore, each sequence should include
(at least once) the symbol 1 (or 4), and the symbol 3 (or 5), otherwise
it should be pruned.

\noindent
(4) An orbit cannot hit a circular boundary, then a straight boundary
and then go back to the same circular boundary. Therefore, symbol
sequences that contain the following segments must be pruned:
...101..., ...104..., ...401..., ...404...,
...121..., ...124..., ...421..., ...424...,
(for the circular boundary on the left), and
...303..., ...305..., ...503..., ...505...,
...323..., ...325..., ...523..., ...525...
(for the circular boundary on the right).

\noindent
(5) An orbit cannot change its orientation between
consecutive hits of a circular boundary.
Therefore, the following sequences are pruned:
...14..., ...41..., ...35..., ...53....

\noindent
(6) An orbit that hits a circular boundary and then follows by
successive hits of the straight boundaries must then hit the second
circular boundary before it can return to the first one
(this is a generalization of pruning rule no 4).
Therefore, sequences such as ...102021..., ...1020204...,
as well as ...3202023..., and ...50205... are geometrically pruned.

We believe that the list presented above contains the complete set
of geometrical pruning rules. Any symbol sequence that falls into
one or more of these six categories will not have a periodic
orbit associated with it no matter what the parameter $a$ is.
We have no proof that this list is complete.
Note that the main use of these rules is to save time by not trying
to calculate orbits that do not exist. If this list is incomplete,
then the extra orbits will be pruned anyway using the dynamical
pruning rules, so in any case it does not affect any of our results.

\figure{
An unstable periodic orbit of order 7 which corresponds to the
symbol sequence 1232131 is shown. Here the length of the straight
boundary is $2 a$ where $a=1.2$, while the radius of the circles
is $R=1$. We use a one dimensional coordinate system
in which $x_n$ represents the position of the nth atom on its
boundary segment. For the straight boundaries $-a<x_n<a$
while for the semicircles  $-\pi/2 <x_n <\pi/2$.
}

\figure{
In this Figure we demonstrate a possible degeneracy that may
occur in a symbolic dynamics of only four symbols.
In this case there may be two different periodic orbits
which correspond to a symbol sequence of the form:
$\dots 0332 \dots$.
To distinguish between them we introduce two more symbols
4 and 5 such that the orbit in (a) for which the direction along
the semicircle is upwards
remains
$\dots 0332 \dots$
while the other one (b) becomes
$\dots 0552 \dots$.
}

\figure{
The artificial dynamics.
In order to reach the minimum of the variational function
the atoms on the straight
boundaries move to the direction of the smaller
angle (a), while the ones on the circular boundaries
move to the direction of the larger angle (b).
We have proved that for the stadium geometry
this choice is correct for all the periodic orbits.
}

\figure{
Most of the periodic orbits in the stadium, such as the one in (a),
are different in phase space from their time reversal.
Therefore, (a) represents two different periodic orbits, which are
symmetric through time reversal, and are both counted. If the
particle is charged and a slight magnetic field is applied
the two orbits will
split and will not coincide anymore even in configuration space.
However, there is a class of periodic orbits, such as the one in
(b), which re-trace their path and are thus equivalent to their time
reversal. These orbits are counted only once.
}

\figure{
The topological entropy as a function of $p$ for the stadium
billiard with $a=1.6$.
}

\figure{
The average discrete Lyapunov exponent
$ \langle \lambda_d \rangle $ given by Eq.
(\ref{liapdis}),
$(\times)$,
and the average continuous Lyapunov exponent
$ \langle \lambda_c \rangle $ given by Eq.
(\ref{liapcon}),
$(\bigtriangleup)$,
as a function of $p$ for the stadium billiard with $a=1.6$.
The horizontal lines represent the values obtained by averaging
over an ensemble of trajectories starting at random initial
conditions. Note that the average Lyapunov exponents for odd $p$ are
significantly larger than for even $p$. Convergence can be improved
by averaging over pairs of consecutive even and odd $p$.
}

\begin{table}
\caption{
The number of unstable periodic orbits
and the topological entropy
of the
stadium billiard for $a=1.6$.
$N_t(p)$ is the total number of symbol sequences of order $p$
which are left after applying the geometric pruning rules
and eliminating cyclic permutations and repetitions of
shorter cycles.
$N_c(p)$ is the number of orbits that remain after the
dynamical pruning rules.
N(p) is the total number of periodic points of order $p$
and its divisors,
and $K_0(p)$ is the $p$th order approximation to the
topological entropy.
}
\begin{tabular}{|r|r|r|r|r|}
\hline
  p   & $N_t(p)$ & $N_c(p)$ & $N(p)$ & $K_0(p)$  \\
\hline
       2 &        1 &        1 &        2 &          0.346\\
       3 &        8 &        8 &       24 &          1.059\\
       4 &       32 &       32 &      130 &          1.216\\
       5 &       96 &       80 &      400 &          1.198\\
       6 &      252 &      188 &     1154 &          1.175\\
       7 &      712 &      376 &     2632 &          1.125\\
       8 &     2096 &     1424 &    11522 &          1.169\\
       9 &     6440 &     4104 &    36960 &          1.168\\
      10 &    19664 &    11064 &   111042 &          1.161\\
      11 &    60488 &    29880 &   328680 &          1.154\\
\hline
\end{tabular}
\end{table}


\begin{references}
\bibitem{twocyc} The only exception is the family of 2-cycles,
                 all associated with the same symbol sequence 02,
                 which bounce between the straight edges.
                 These are the only periodic orbits in the stadium
                 which are not unstable and not isolated.
\bibitem{directions} By considering the derivatives
                 $d \theta_n^{\pm}/d x_n$ as a function of
                 $x_n$ for the stadium geometry, one can prove
                 that the indicated choice of sign for
                 $C_n$ is correct for all orbits.
\bibitem{shrt} Techniques which are based on map iteration
               cannot provide such a systematic calculation.
               First, one cannot guarantee
               that all the orbits up to a given order have
               been found.
               Second, the exponential sensitivity to initial
               conditions
               and the finite precision of the computer
               set an upper bound on the length of unstable periodic
               cycles which can
               be calculated.
\bibitem{nopo} Note that the number of periodic points
               which belong to cycles of order $p$ and its divisors,
               is
               $N(p)=\sum i \cdot N_c(i)$
               where $N_c(i)$ is the number of periodic cycles of
               order $i$ and
               the sum is over all the divisors of
               $p$ including $p$.
\end{references}
\end{document}